\documentclass[aps,prl,twocolumn,groupedaddress]{revtex4-1}
\usepackage{graphicx}
\usepackage{dcolumn}
\usepackage{bm}
\usepackage{amsmath}
\usepackage{amssymb}
\usepackage{latexsym}
\usepackage{epsfig}
\usepackage{amsbsy}
\usepackage{array}
\usepackage{amssymb}
\usepackage{setspace}
\usepackage{bm}
\usepackage{xcolor}
\usepackage{multirow}

\def\sint{\ifmmode{- \!\!\!\!\!\! \int}
    \else{\hbox{$- \!\!\!\! \int \ $}}\fi}

\begin{document}

\title{Topological Larkin-Ovchinnikov phase and Majorana zero mode chain in bilayer superconducting topological
insulator films}


\author{Lun-Hui Hu$^{1,2,3,4}$}
\author{Chao-Xing Liu$^{3}$}
\email{cxl56@psu.edu}
\author{Fu-Chun Zhang$^{1,4}$}
\email{fuchun@ucas.ac.cn}

\affiliation{$^1$Kavli Institute for Theoretical Sciences,  University of Chinese Academy of Sciences, Beijing 100190, China}
\affiliation{$^2$Department of Physics, Zhejiang University, Hangzhou, Zhejiang, 310027, China}
\affiliation{$^3$Department of Physics, The Pennsylvania State University, University Park, Pennsylvania 16802, USA}
\affiliation{$^4$Collaborative Innovation Center of Advanced Microstructures, Nanjing 210093, China}

\date{\today}

\begin{abstract}
We theoretically study bilayer superconducting topological insulator film, in which superconductivity exists for both top and bottom surface states. We show that an in-plane magnetic field
can drive the system into Larkin-Ovchinnikov (LO) phase, where electrons are paired with finite momenta.
The LO phase is topologically non-trivial and characterized by a $\mathbb{Z}_2$ topological invariant, leading
to a Majorana zero mode chain along the edge perpendicular to in-plane magnetic fields.
\end{abstract}


\maketitle

{\it{ Introduction.-}}
Magnetism and superconductivity are two fundamental states of matter in condensed matter physics
and the interplay between them continues bringing us intriguing phenomena. Unlike the conventional
Cooper pairs with zero momentum in the BCS theory, magnetism can induce a superconducting (SCing) state
with finite momentum pairing. The pairing function of such state can either carry a single finite momentum $Q$, known
as Fulde-Ferrell (FF) state\cite{fulde_pr_1964}, or multiple finite momenta $Q_i$ ($i=1,2,\dots$), known as
Larkin-Ovchinnikov (LO) state\cite{larkin_jetp_1965}. There are extensive experimental efforts aiming in
realizing FF or LO states
in various systems, including heavy fermion superconductors (SCs), cold atom systems and  organic
SCs\cite{cr_rmp_2004,matsuda_jpsj_2007,lr_rpp_2010}.
Another recent development is to realize topological SCs by integrating magnetism, spin-orbit coupling and superconductivity into one hybrid system \cite{nayak_rmp_2008,alicea_rpp_2012,elliott_rmp_2015,sato_rpp_2017,qi_prb_2010},
in which gapless excitations exist at the boundary or in the vortex core, dubbed ``Majorana fermions''
or ``Majorana zero mode (MZM)''. MZM possess exotic non-Abelian statistics and thus can serve as the building
block for topological quantum computation \cite{kitaev_pu_2001,kitaev_aop_2003}.

Since both finite momentum pairing and topological superconductivity require magnetism
and superconductivity, it is natural to ask if these two SCing phenomena can coexist and if there is
any interplay between them.
In particular, one may ask (1) if topological SC phases
can exist for FF or LO state with finite momentum pairing;
(2) how to find an experimentally feasible system for a robust realization of such state; and
(3) what types of boundary modes can emerge in such system.

In this work, we propose a new topological LO (tLO) state in topological insulator (TI)
thin films in proximity to conventional s-wave SCs
under an in-plane magnetic field. By combining a general theoretical argument of topological invariant,
self-consistent calculation of phase diagram and the direct calculation of edge modes, we demonstrate the
existence of tLO phase in this bilayer SCing TI films in a wide
parameter regime. In particular, we show a chain of numerous MZMs, dubbed ``MZM chain'',
existing along the 1D edge perpendicular to the in-plane magnetic field.

{\it Topological LO state.-}
We start from a general discussion of the possibility of topological nature of LO state.
For a SCing system, we consider the Bogoliubov-de Gennes (BdG) Hamiltonian
with the single particle Hamiltonian $\mathcal{H}_0(k_x,-i\partial_y)$ and the gap function
$\Delta(y)$. The gap function satisfies periodic condition $\Delta(y)=\Delta(y+2\pi/Q)$
with the wavevector $Q$ and can be expanded as $\Delta(y)=\sum_{n}\Delta_n e^{inQy}$
with an integer $n$.
Only one $\Delta_n$ is non-zero in the FF state while multiple non-zero $\Delta_n$ exist
in the LO state.
The BdG Hamiltonian takes the form
\begin{eqnarray}
	\mathcal{H}_{BdG}=\left(
	\begin{array}{cc}
		\mathcal{H}_{ee} & \mathcal{H}_{eh}\\
		\mathcal{H}_{he} & \mathcal{H}_{hh}
	\end{array}
	\right),\label{Eq:HamBdG}
\end{eqnarray}
where $\mathcal{H}_{ee}$, $\mathcal{H}_{hh}$ are Hamiltonians for electrons and holes, respectively,
and $\mathcal{H}_{eh}$ represents paring. $\mathcal{H}_{BdG}$ may be expanded in the momentum space,
leading to the form
$\left(\mathcal{H}_{ee}\right)_{nm}=\delta_{nm} \mathcal{H}_0(k_x,nQ+k_y)$, $\left(\mathcal{H}_{hh}\right)_{nm}=-\delta_{nm}\mathcal{H}^{*}_0(-k_x,nQ-k_y)$,
$\left(\mathcal{H}_{eh}\right)_{nm}=\Delta_{n+m}$ and $\left(\mathcal{H}_{he}\right)_{mn}=\Delta_{n+m}^{\dag}$ ($n,m$ are integer numbers)
on the basis $\left\vert e_n \right\rangle$ and $\left\vert h_n\right\rangle$ with the wavevector $nQ$.
Here the momentum $k_y$ is within the reduced Brillouin zone $[0,Q]$.
The Hamiltonian \eqref{Eq:HamBdG} possess particle-hole symmetry $\mathcal{C}=t_x \mathcal{K}$ where the Pauli matrix
$t_x$ acts on the particle-hole space and $\mathcal{K}$ is the complex conjugate.
Based on the Hamiltonian \eqref{Eq:HamBdG}, we can extract the topological property of
FF state, as shown in Supplementary materials\cite{supp_mat}, which is consistent with
the recent results on topological FF state in cold atom systems
\cite{qu_nc_2013,zhang_nc_2013,wu_prl_2013,cao_prl_2014}.

Next we focus on the LO state with non-zero $\Delta_{\pm 1}$, for which
the Hamiltonian \eqref{Eq:HamBdG} can be split into two decoupled blocks. All the
electron part $\mathcal{H}_0(k_x,nQ+k_y)$ with even (odd) $n$ is only coupled to the hole part $-\mathcal{H}^\ast_0(-k_x,nQ-k_y)$
with odd (even) $n$. We call these two blocks as even and odd block, denoted as $\mathcal{H}_{BdG}^{even}$
and $\mathcal{H}_{BdG}^{odd}$, respectively. Here the even block is written on the basis $\left\vert e_{2n} \right\rangle$
and $\left\vert h_{2n+1} \right\rangle$, while the odd block is written on the basis $\left\vert e_{2n-1}\right\rangle$
and $\left\vert h_{2n} \right\rangle$.
The global particle-hole symmetry $\mathcal{C}$ relates these two blocks,
$\mathcal{C} \mathcal{H}_{BdG}^{even} \mathcal{C}^{-1}=-\mathcal{H}_{BdG}^{odd}$, and thus there is in general no particle-hole symmetry within one block.
However, at the momentum $k_y=Q/2$, a new particle-hole symmetry operator $\tilde{\mathcal{C}}$
can be defined as $\tilde{\mathcal{C}}\left\vert e_{2n}\right\rangle=\left\vert h_{2n+1}\right\rangle$ and $\tilde{\mathcal{C}}\left\vert h_{2n+1}\right\rangle
=\left\vert e_{2n}\right\rangle$ for the even block $\mathcal{H}_{BdG}^{even}$ and
$\tilde{\mathcal{C}}\left\vert e_{2n-1}\right\rangle=\left\vert h_{2n}\right\rangle$ and $\tilde{\mathcal{C}}\left\vert h_{2n}\right\rangle
=\left\vert e_{2n-1}\right\rangle$ for the odd block $\mathcal{H}_{BdG}^{odd}$, and we have
$\tilde{\mathcal{C}}\mathcal{H}_{BdG}^{even}(k_x,Q/2)\tilde{\mathcal{C}}^{-1} = -\mathcal{H}_{BdG}^{even}(-k_x,Q/2)$ and
$\tilde{\mathcal{C}}\mathcal{H}_{BdG}^{odd}(k_x,Q/2) \tilde{\mathcal{C}}^{-1} = -\mathcal{H}_{BdG}^{odd}(-k_x,Q/2)$,
as shown in supplementary materials\cite{supp_mat}. The existence of this new particle-hole symmetry $\tilde{\mathcal{C}}$
suggests that the Hamiltonian $\mathcal{H}_{BdG}^{even}$ and $\mathcal{H}_{BdG}^{odd}$ can be viewed as a one dimensional (1D)
SC chain in the D class along the x direction at $k_y=Q/2$.
A $\mathbb{Z}_2$ topological invariant \cite{kitaev_pu_2001} $\mathcal{M} = \text{sgn}\left( \text{Pf}\left\lbrack A(k_x=0)\right\rbrack
\text{Pf}\left\lbrack A(k_x=\pi)\right\rbrack \right)$ can be defined, where the anti-unitary matrix $A$ is
the BdG Hamiltonian $\mathcal{H}_{BdG}^{even}$ or $\mathcal{H}_{BdG}^{odd}$ at $k_y=\frac{Q}{2}$ in the Majorana representation. Here $\mathcal{M}=-1$ is for topologically non-trivial phase with a pair of MZM present, while
$\mathcal{M}=+1$ is for trivial phase. Thus we conclude that
tLO state is possible to exist with $\mathbb{Z}_2$ classification due to
the new particle-hole symmetry $\tilde{\mathcal{C}}$ that is only valid at $k_y=Q/2$.
Below we discuss how
to realize tLO state in a SCing TI film and the corresponding gapless boundary modes, i.e., MZM chain.

\begin{figure}[t]
  \centering
  \includegraphics[width=3.375in]{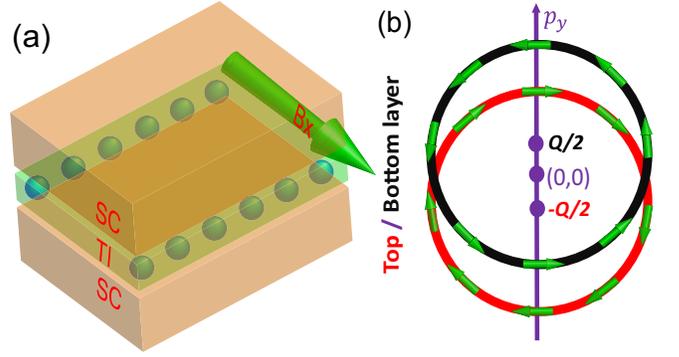}
  \caption{\label{fig-sketch}(Color online) (a) Illustration of the SC/TI/SC heterostructure under an in-plane magnetic field $B_x$ possessing MZM chain at the boundary, and (b) Illustration of the two Fermi surfaces with opposite spin textures
  for TI films with $m=0$ in Eq. (\ref{eq-ham-continue}) under an in-plane magnetic field $B_x$.}
\end{figure}

{\it Bilayer SCing TI film.-}
Here we consider a SCing TI film with both top and bottom surface states (bilayer)
under an in-plane magnetic field, as shown in Fig.~\ref{fig-sketch}.
The low energy physics of this system can be described by the Hamiltonian,
\begin{align}\label{eq-ham-total}
\mathcal{H}=\mathcal{H}_0+\mathcal{H}_{pair}.
\end{align}
Here $\mathcal{H}_0$ describes surface states at the top and bottom surfaces under an in-plane magnetic field
along the x direction and is given by
\cite{lu_prb_2010,yu_science_2010}
\begin{align}\label{eq-ham-continue}
  \mathcal{H}_0 = \int d\mathbf{r}\, \tilde{c}^\dagger(\mathbf{r}) \left\lbrack m\tau_x +
  v\tau_z(\hat{p}_x \sigma_y - \hat{p}_y \sigma_x) + B_x\sigma_x\right\rbrack \tilde{c}(\mathbf{r})
\end{align}
where $\tilde{c}=\left(\hat{c}_{t,\uparrow},\hat{c}_{t,\downarrow},\hat{c}_{b,\uparrow},\hat{c}_{b,\downarrow}\right)^\text{T}$ are the electron annihilation operators, $(\hat{p}_x, \hat{p}_y)$ are in-plane momentum operators, $v$ is the Fermi velocity, and $\sigma$ and $\tau$ are Pauli matrices,
representing spin and pseudo-spin (top and bottom surfaces), respectively.
For the Hamiltonian (\ref{eq-ham-continue}),
the first term ($m = m_0 + m_1 \hat{\mathbf{p}}^2$) describes
the tunneling bewteen two surface states
(called inter-layer tunneling below),
the second term is the Dirac Hamiltonian for two surface states,
and the third term gives
the Zeeman coupling between electron spin and in-plane magnetic fields.
Here we have absorbed the parameters $\hbar$ into $v$ and $g\mu_B$ into $B_x$
to simplify the notation.
To include the SC pairing, we consider on-site electron-electron
attractive interaction term within one surface state
(called intra-layer below) as
\begin{eqnarray}\label{eq-ham-pair}
  \mathcal{H}_{pair} = -U\int d\mathbf{r}\sum_{\tau} \; \hat{n}_{\tau,\uparrow}(\mathbf{r}) \hat{n}_{\tau,\downarrow}(\mathbf{r})
\end{eqnarray}
where $\hat{n}_{\tau,\sigma}(\mathbf{r})$ is the electron density operator of spin $\sigma=\uparrow, \downarrow$
on layer $\tau=t,b$. Here we neglect the inter-layer interaction, which should be repulsive and weaker.

\begin{figure}[!htbp]
  \centering
  \includegraphics[width=3.37in]{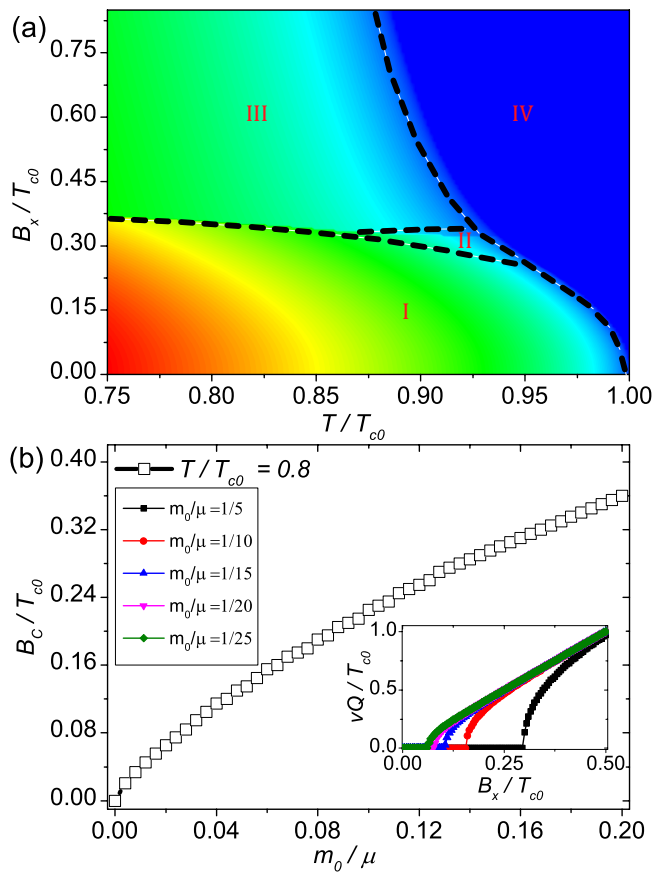}
  \caption{\label{fig-phase-diagram}(color online) (a) Phase diagram in parameters of $B_x/T_{c,0}$ and $T/T_{c,0}$ at $m_0/\mu=1/5$ for intra-layer pairing, comprising conventional BCS phase (I), FF phase (II), LO phase(III) and normal metal (IV). (b) The critical field $B_c/T_{c,0}$ (transition between BCS to LO phase) as function of $m_0$ at fixed $T/T_{c,0}=0.8$. Inset shows $\nu Q/T_{c,0}$ as function of $B_x/T_{c,0}$ for different $m_0/\mu$.  Parameters used here are $m_1=0$, $UN_0=0.15$, $\omega_D=100$ meV and $\mu=100$ meV, so that $T_{c,0}=1.134\omega_D \exp(-1/UN_0)=0.1443$ meV.}
\end{figure}

The phase diagram in Fig.~\ref{fig-phase-diagram}(a) can be constructed by minimizing Ginzburg-Landau free energy $\mathcal{L}$
obtained from the microscopic Hamiltonian \eqref{eq-ham-total}\cite{sigrist_rmp_1991}. The details of the calculation for the phase diagram are presented
in the supplementary material\cite{supp_mat}.
The gap function takes a general form
$\Delta_t\left( \frac{1+\tau_z}{2} \right)
\sigma_y +\Delta_b\left( \frac{1-\tau_z}{2} \right)\sigma_y$,
where $\Delta_t$ and $\Delta_b$ are the gap functions for the top and bottom surfaces, respectively.
At $B_x=0$, the BCS type of intra-layer spin-singlet pairing with
$\Delta_t=\Delta_b=\Delta_0$ will be energetically favored.
However, when increasing $B_x$, the momentum $Q$ of the pairing
starts preferring some non-zero value at a critical magnetic field
$B_c$, as shown in the inset of Fig.~\ref{fig-phase-diagram}(b),
suggesting a phase transition occurring.
The phase diagram as a function of temperature $T$ and
in-plane magnetic field $B_x$ is shown in Fig.~\ref{fig-phase-diagram}(a)
for $m_0/\mu=1/5$.
Four phases, including (I) BCS state ($\Delta_0\sigma_y$),
(II) FF state ($\Delta_t=\Delta_0e^{-iQy}, \Delta_b=0$ or $\Delta_t=0, \Delta_b=\Delta_0e^{iQy}$),
(III) LO state ($\Delta_t=\Delta_b^\ast=\Delta_0e^{-iQy}$),
and (IV) normal metallic state, are identified.
At low temperatures, the BCS state is favored for a small $B_x$,
while the LO state is present for a large $B_x$.
Near the transition temperature between SCing states and normal state,
we find the FF state existing in a small region between the BCS and LO states.
However, this small region for FF state will disappear
for a smaller $m_0$ [See supplementary materials\cite{supp_mat}].
The transition between the BCS pairing and the FF or LO state is of first order
and occurs at a critical magnetic field strength $B_{c}$
along the transition temperature line between SCing states and normal state
In the inset of Fig.~\ref{fig-phase-diagram}(b),
different color lines are for different $m_0/\mu$ and thus the
critical $B_c$ depends on the coupling ratio $m_0/\mu$.
From Fig.~\ref{fig-phase-diagram}(b), we notice that
the value of $B_c$ approaches zero when turning off $m_0$.

To understand the occurrence of LO state,
we may first consider the energy spectrum
of the single-particle Hamiltonian $\mathcal{H}_0$ in Eq.~\eqref{eq-ham-continue}, which is given by
$E_{0,s}(\mathbf{p})=\pm \sqrt{v^2p_x^2 + \left(\sqrt{m^2+v^2p_y^2} + s B_x\right)^2}$,
with $s=\pm 1$.
In the decoupling limit ($m\rightarrow0$),
the Fermi surfaces of two surface states are shifted with $\pm Q/2$
in the opposite directions with $vQ=2B_x$
due to the Zeeman term, as illustrated in Fig.~\ref{fig-sketch}(b).
Spin textures of the surface states are also depicted on the Fermi surfaces, from which one can see
that zero momentum pairing can only occur for electrons with the same spin
(equal spin triplet pairing), while spin-singlet pairing
is only possible for a finite momentum.
In the limit $m\rightarrow 0$, the LO phase is always favored and
can be viewed as two FF phases with opposite momenta
for each surface state,
similar to the LO phase in the bilayer TMDs system\cite{liu_prl_2017}.
A finite coupling term $m$ will induce
a Josephson coupling between $\Delta_t$ and $\Delta_b$ pairing,
which tends to induce a bonding state ($\Delta_t=\Delta_b=\Delta_0$)
in order to lower the free energy $\mathcal{L}$. However, the opposite momentum shift will make such
Josephson coupling vanishing and as a result, there is a competition between Josephson coupling
due to the finite $m$, which favors BCS pairing, and the momentum shift due to in-plane magnetic fields,
which favors FF or LO state. Thus, a finite $m$ will increase the critical magnetic field $B_c$ [See Fig.~\ref{fig-phase-diagram}(b)].
We notice that our phase diagram (Fig. \ref{fig-phase-diagram}(a)) is quite similar to
that of 2D Rashba SCs \cite{barzykin_prl_2002,dimitrova_prb_2007} due to the same
spin textures (Fig.~\ref{fig-sketch}(b)) in these two systems.

{\it Majorana zero mode chain.-}
We next focus on topological properties of the LO state found in the last section.
For the convenience of calculations, we consider the lattice regularization of
the BdG Hamiltonian \eqref{eq-ham-total}.
The lattice version of the single-particle Hamiltonian $\mathcal{H}_0$ reads
\begin{equation}\label{eq_TB_Ham}
  \begin{split}
    \mathcal{H}_0 &= \sum_{i_x}\sum_{i_y} \Big{\{} \tilde{c}^\dagger(i_x,i_y) \left( T_0 \right) \tilde{c}(i_x,i_y) \\
     &\quad + \tilde{c}^\dagger(i_x,i_y) \left( T_x \right) \tilde{c}(i_x+1,i_y) \\
     &\quad + \tilde{c}^\dagger(i_x,i_y) \left( T_y \right) \tilde{c}(i_x,i_y+1) + \text{H.c.}  \Big{\}}
  \end{split}
\end{equation}
where the integers $i_x$ and $i_y$ describes the lattice sites (the lattice constant is chosen to be 1),
$T_0 = (m_0+4m_1)\tau_x + B_x \sigma_x$, $T_x = i\alpha \tau_z\sigma_y - m_1\tau_x$, and $T_y = -i\alpha \tau_z\sigma_x - m_1\tau_x$ with $\alpha=v/2$. In the BdG Hamiltonian, we set $\mathcal{H}_{ee}=\mathcal{H}_0$ and $\mathcal{H}_{hh}=-\mathcal{H}_0^*$.
For the LO state, the off-diagonal part of $\mathcal{H}_{BdG}$ is given by
$\mathcal{H}_{eh}=\Delta_0e^{iQy}\left( \frac{1+\tau_z}{2} \right)
\sigma_y +\Delta_0e^{-iQy}\left( \frac{1-\tau_z}{2} \right)\sigma_y$.
Here we always choose $Q=2\pi/N_y$ with an integer $N_y$ for convenience.
Due to the periodicity of the BdG Hamiltonian
$\mathcal{H}_{BdG}(i_x,i_y)=\mathcal{H}_{BdG}(i_x,i_y+N_y)$,
the parameter $N_y$ gives the y-directional length of the supercell.
Since $Q$ depends on the magnetic field $B_x$ ($Q=2B_x/v$ for
$m\rightarrow 0$ or large enough $B_x$),
the length $N_y$ of the supercell also depends on $B_x$ and will be reduced
when $B_x$ is increased. Motivated by our general theory for tLO phase,
we next study the energy dispersion of a slab configuration for the BdG Hamiltonian
which is finite along the x direction ($N_x$ sites) and infinite along the y direction.
According to the Bloch theorem, we need to solve the eigen-equation
$\mathcal{H}_{BdG}\vert\psi\rangle=E\vert\psi\rangle$ in a super-cell of $N_x\times N_y$ lattice sites with open boundary
condition along the x direction and twist boundary condition $\left\vert \psi(i_x,N_y)\right\rangle=e^{ik_yN_y}\left\vert \psi(i_x,0)\right\rangle$
along the y direction for any $i_x=1,\dots,N_x$ ($k_y\in[0,Q]$).

\begin{figure}[t]
  \centering
  \includegraphics[width=3.3in]{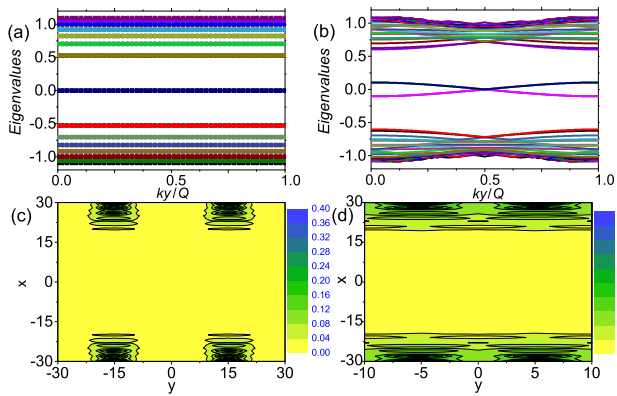}
  \caption{\label{fig-mzm-pure-fflo}(color online) Low eigenenergy spectrum (a,b) and probability distribution of MZM state (c,d) in the LO phase with \(\Delta_t=\Delta_0 e^{-iQy}\) and \(\Delta_b=\Delta_0 e^{iQy}\). Open boundary along x axis and periodic boundary along y axis are used.
  In (a)(c), $Q=2\pi/61$ and the unit cell size is $N_x=N_y=61$; MZM chain appears with a large minigap $\Delta E\sim 0.4\Delta_0$.
  In (b)(d), $Q=2\pi/21$ and the unit cell size is $N_x=61, N_y=21$; Majorana bands disperse due to the hybridization between the intra-edge MZMs, however two pair of MZMs at $k_y=Q/2$ are protected.
  Parameters used here are $\alpha=m_1=5$, $m_0=0$, $B_x=2\alpha\sin(Q)$, $\Delta_0=1.5$ and $\mu=13.5$.}
\end{figure}

The energy dispersions of the slab are shown in
Fig.~\ref{fig-mzm-pure-fflo}(a) for $N_x=N_y=61$
and (b) for $N_x=61, N_y=21$.
For a large $N_y$ (corresponding to a small $B_x$),
flat bands are found at zero energy ($E_0\sim\pm10^{-6}$) in Fig.~\ref{fig-mzm-pure-fflo}(a),
suggesting the existence of highly localized MZMs.
The local density of states in a super-cell for flat Majorana bands are shown in Fig.~\ref{fig-mzm-pure-fflo}(c),
from which one indeed finds two pairs of MZMs located at two edges of the slab.
According to our numerical simulations, we find that the MZMs are located
at $y=\pi/2Q + n\pi/Q$ ($n$ is an integer) with the localization length
estimated as
$\xi\sim v/\Delta_0 \sim 6$, which is much smaller compared to the length
$N_y=61$ of the supercell.
Thus, these MZMs are well separated and form the flat Majorana bands.
In addition to the flat Majorana bands, there are also topologically trivial Andreev bound states within the SCing gap.
These bound states are well separated from MZMs with an energy gap $\Delta E\sim0.4\Delta_0$.

The above analysis also indicates that adjacent MZMs might be hybridized
when $N_y$ is reduced. Indeed, for a small $N_y=21$ (corresponding to a large $B_x$),
we find the Majorana bands become dispersive,
as shown in Fig.~\ref{fig-mzm-pure-fflo}(b).
The strong hybridization between
MZMs at one edge is shown in Fig.~\ref{fig-mzm-pure-fflo}(d).
We notice that these two Majorana
bands cross with each other at $k_y=Q/2$. This crossing with four-fold degeneracy can be explained by the
new particle-hole symmetry $\tilde{\mathcal{C}}$ defined at $k_y=Q/2$,
which is consistent with the general theory for tLO phase discussed above.
Thus, our calculation demonstrates the $\mathbb{Z}_2$ tLO phase can indeed be realized in our bilayer SCing TI films.

{\it Discussion and conclusion -}
In this work, we develop a general theory of tLO phase with $\mathbb{Z}_2$ classification
and propose its material realization in bilayer SCing TI films.
The realization of tLO phase and the corresponding MZM chain
opens a new route in the study of MZMs for quantum computation.
The 1D MZM chain also provides a natural platform to stuy
interacting Majorana chains\cite{rahmani_prl_2015,chiu_prb_2015}.

The proposed model can be realized in SC/TI/SC hetero-structure
\cite{xu_prl_2014,xu_prl_2015,sun_prl_2016}, e.g. NbSe$_2$/Bi$_2$Te$_3$/NbSe$_2$ heterostructure.
With the parameters $\Delta_0=1$ meV, $\mu=100$ meV, $\hbar v=0.4$ nm$\cdot$eV and
the g-factor $g\approx20$\cite{wang_prb_2011}, and $m_0=\mu/5$, we can estimate the critical field
at tricritical point is about $0.17$ Tesla according to $g\mu_BB_c/k_BT_{c,0}\approx 0.35$ from Fig.~\ref{fig-phase-diagram}(b).
The distance between two MZMs is estimated as $\Delta_y= \pi \hbar v/4g\mu_B B_c\approx 1.6$ $\mu$m,
which is four times larger than the localization length of MZMs $\xi \sim \hbar v/\Delta_0=0.4$ $\mu$m.
Thus, MZMs in the chain are well separated and can be resolved in a scanning tunneling
microscope experiment \cite{sun_prl_2016}.
The above estimate is based on Zeeman effect, but we emphasize that
the orbital effect of in-plane magnetic fields can also plays a similar role as the Zeeman effect
due to the Dirac fermion nature. Compared to the Zeeman effect, we find the orbital effect of an in-plane magnetic field can also induce the FFLO phase (the vector potential could be chosen as $\vec A=(0,-B_x z,0)$ and set the middle of layers as $z=0$ resulting in the opposite momentum shift for the Fermi surface of top and bottom surface states), and it is about $\pi\hbar v d/2\Phi_0 \sim 1.21$ meV/Tesla by assuming the space distance between two surfaces $d=4$ nm, which is comparable to the Zeeman term with $g\mu_B=1.16$ meV/Tesla.
One can also consider SC/magnetic TI/SC heterostructure, in which the exchange coupling from
magnetic doping takes a similar form as Zeeman effect, but is two orders of magnitude larger than the Zeeman effect
\cite{yu_science_2010}.
Based on the above estimate, we conclude that our proposal is feasible under the current experimental conditions.

Our proposal is also applicable to SCing TIs in which topological surface states
and bulk superconductivity can coexist. Such materials include Cu doped Bi$_2$Se$_3$
\cite{wray_nat_phy_2010,hor_prl_2010,hor_prl_2010,kriener_prl_2011}, several SCing
half-Heusler compounds (e.g. YPtBi, RPdBi) \cite{lin_nat_mat_2010,nakajima_sci_adv_2015}
and FeTe$_{0.55}$Se$_{0.45}$ \cite{yin_nat_phys_2015,zhang_arxiv_2017}.

{\it Acknowledgement -}
It is a pleasure to thank Cheung Chan, Chuang Li, James Jun He, Jia-Bin Yu, Jian-Xiao Zhang and Rui-Xing Zhang for the helpful discussions.
C.-X.L. acknowledge the support from Office of Naval Research (Grant No. N00014-15-1-2675). FCZ is partly supported by NSFC grant 11674278 and National Basic Research Program of China (No. 2014CB921203).

\bibliographystyle{apsrev4-1}
\bibliography{Reference_main}

\end{document}